\documentclass[12pt,preprintnumbers,amsmath,amssymb,nofootinbib]{revtex4-1}
\usepackage{graphicx}
\usepackage{amsmath}
\usepackage{amssymb}
\usepackage{bm}
\usepackage{multirow}
\usepackage{natbib}

\def\Im{\text{Im}}

\begin{document}

\title{Nature of X(3872) from recent BESIII data: Considering the universal feature of an S-wave threshold resonance}

\author{Xian-Wei Kang}
\author{Jin-Zhe Zhang}
\author{Xin-Heng Guo}

\affiliation{Key Laboratory of Beam Technology of Ministry of Education, School of Physics and Astronomy, Beijing Normal University, Beijing 100875, China\\
and Institute of Radiation Technology,
Beijing Academy of Science and Technology, Beijing 100875, China}

\begin{abstract}
We analyze the recent data from the BESIII collaboration on the $X(3872)$ state in the $J/\psi\pi^+\pi^-$ and $D^0\bar{D}^0\pi^0$ decay channels. The quantum number and mass of the $X(3872)$ state allow us to exploit the universal feature of the very near-threshold $D\bar D^*$ scattering in the $S$ wave. The analysis of $J/\psi\pi^+\pi^-$ data and $D^0\bar{D}^0\pi^0$ data separately as well as the combined analysis of these data together, all support the conclusion that $X(3872)$ is an extremely weakly bound charm meson
molecule.
\end{abstract}

\maketitle

\section{INTRODUCTION}
The meson (quark antiquark) and baryon (three quarks) constitute the traditional configuration of a hadron. Whether other configurations exist, e.g., the multiquark state or hybrid state of multiquarks and gluons, remains an interesting question all along. Since the discovery of $X(3872)$ by the Belle collaboration in 2003 \cite{Belle:2003nnu}, many such exotic $XYZ$ states have been detected in experiments. Their structures are difficult to accommodate with simple quark models and thus arouse great interest in their inner structures in our community.
%Quality Control Editor: Please ensure that the intended meaning has been maintained in the following edit.
Tremendous progress has been made in the past two decades (e.g., see reviews: \cite{Guo:2017jvc,Brambilla:2019esw,Chen:2022asf,Meng:2022ozq,Liu:2024uxn})
. It is now mostly regarded as a tetraquark state or a molecule. In Ref.~\cite{Kang:2016jxw}, it was shown that the invariant mass spectrum data are not enough to determine the inner structure of
$X(3872)$, and thus, its properties also need to be analyzed by considering various decay patterns. There are recent studies concerning strong decay \cite{Wang:2023sii,Achasov:2024ezv,Dias:2024zfh} and radiative decay \cite{Yu:2023nxk,BESIII:2024ync}. In 2023, BESIII published their
data on $J/\psi\pi^+\pi^-$ and $D^0\bar{D}^0\pi^0$ data in the region of $X(3872)$ \cite{BESIII:2023hml}. The new data prompted us to reexamine the existing analysis of $X(3872)$.

The quantum number of $X(3872)$ is known to be $J^{PC} = 1^{++}$ \cite{ParticleDataGroup:2024cfk}, with $J$, $P$, and $C$ representing the spin, parity and charge conjugation, respectively. Its mass is $3871.64\pm0.06$ MeV \cite{ParticleDataGroup:2024cfk}. Its extreme proximity to the $D^0  \bar D^{*0}$ mass threshold ($m_{\rm th}=$3871.69 MeV) suffices to consider only the $S$-wave $D\bar D^*$ scattering, where $D\bar D^*$ actually refers to the $C=+$ combination of $(D^0\bar D^{*0}+\bar D^0 D^{*0})/\sqrt{2}$. We will adopt the universal assumption to treat the low-energy
$D\bar D^*$ scattering, in which the scattering length dominates over all other terms in the expansion of the amplitude \cite{Braaten:2007dw,Braaten:2007ft,Braaten:2009jke,Song:2023pdq}. In this way, we will examine whether the simpler method used in Ref.~\cite{Braaten:2009jke} can work for the new BESIII data on $X(3872)$ \cite{BESIII:2023hml}. And especially, this new data set concerns the
mass distribution of $D^0\bar D^0\pi^0$ but not for the $D^0\bar D^{*0}$ as in previous analysis \cite{Braaten:2009jke}.
This constitutes for our main motivation and theoretical framework. Note that the accidental
nearness of the mass values between $m_{D^*}-m_D$ and $m_\pi$ may invalidate the perturbative treatment of pions. However, the authors in \cite{Fleming:2007rp} demonstrated that the much weaker $DD^*\pi$ (vertex compared to the $NN\pi$ one) together makes the perturbation theory work.
In addition, we also include the effect of the finite $D^{*0}$ width due to the small binding energy. To consider the inelastic effect, e.g., the scattering between $J/\psi\pi^+\pi^-$ and $D^0 \bar D^{0*}$, the scattering length is extended to be complex-valued. This treatment has been extensively used and has also been verified to work successfully, e.g., in processes involving $p\bar p$, where the annihilation effect is strong \cite{Kang:2013uia,Kang:2015yka,Haidenbauer:2015yka}. Then, the formalism of energy-dependent event distribution can be established from the $D^0 \bar D^{0*}$ scattering amplitude with the scattering length as the only free parameter. By fitting to the new BESIII data \cite{BESIII:2023hml}, we can fix the parameter and deduce the relevant physical information behind it. The interpretation of $X(3872)$ as the $D^0 \bar D^{0*}$ bound state (or as the charm meson molecule) is consistent with the new BESIII data.

The paper is organized as follows: After this introduction, we present the derivation of our formalism for the event distribution for the $J/\psi\pi^+\pi^-$ and $D^0\bar{D}^0\pi^0$ processes in Sec.~II. In Sec.~III, we present the fit results and our discussion, where both the separate fit and the simultaneous fit are performed. We close with a summary in Sec.~IV.

\section{Formalism: the scattering amplitude and decay rate}
As mentioned above, we consider the universal feature of the near-threshold $D^0 \bar D^{0*}$ scattering in the $S$ wave. In nonrelativistic quantum mechanics, the amplitude for pure elastic scattering between two stable particles can be written as
\begin{equation}\label{Eq:fEelastic}
f(E)=\frac{1}{-\gamma+\kappa(E)},
\end{equation}
where $\gamma=1/a$ is the inverse scattering length, $\mu=966.6$ MeV is the reduced mass of $D^{*0}$ and $\bar{D}^0$, and $\kappa(E)=\sqrt{-2\mu E-i\epsilon}$ with $\epsilon=0^+$. Here, $E$ is defined as the total energy of $D^0 \bar D^{0*}$ relative to their mass threshold. For $E<0$, $\kappa(E)$ is a real and positive number. For $E>0$, $\kappa(E)$ takes the value of $-ik$ with $k>0$. The correct choice of the sign of $i\epsilon$ ensures this point, and Eq.~\eqref{Eq:fEelastic} agrees with the common scattering length approximation. Eq.~\eqref{Eq:fEelastic} fulfills the unitarity constraint for the single-channel system provided that $\gamma$ is real-valued:
\begin{equation}\label{eq:uni-elas}
{\Im} f(E)=|f(E)|^2 \sqrt{2\mu E}, \qquad E>0,
\end{equation}
which is just the optical theorem up to some kinematic factors.
For the case of $E<0$, i.e., below the two-body threshold, one will encounter a pole in $f(E)$: we stress that a bound state occurs if $\gamma>0$ and a virtual state occurs if $\gamma<0$. The former is found in the first Reimman sheet, and the latter is found in the second sheet.

To consider the inelastic scattering between $D^0 \bar D^{0*}$ and $J/\psi\pi^+\pi^-$, the value of $\gamma$ is taken to be complex:
i.e., $\gamma=\gamma_{\rm re}+i \gamma_{\rm im}$. A similar treatment has been performed for proton--antiproton scattering \cite{Kang:2013uia,Kang:2015yka,Haidenbauer:2015yka}. In this case, the imaginary part of the amplitude becomes
\begin{equation}\label{eq:uni-multi}
\Im f(E)=|f(E)|^2(\gamma_{\rm im}-\Im\kappa(E)).
\end{equation}
This can be understood as the unitarity fulfilled by the multichannel system. The imaginary part of $\gamma$, $\gamma_{\rm im}$, is proportional to
the inelastic cross-section, and thus $\gamma_{\rm im}>0$. By a more explicit derivation, we note the unitarity condition \footnote{The $S$-matrix element is expressed by $S(E)=1+2i f(E)$, and $SS^\dagger\leq 1$ leads to Eq.~\eqref{eq:1311}. The $a(\boldmath{q})$ in Ref.~\cite{Kang:2013uia} plays the role of $f(E)$ here. For a practical calculation, e.g., the $p\bar p$ potential in chiral effective field theory, the imaginary part of the scattering length in Refs.~\cite{Kang:2013uia,Kang:2015yka} is negative for all, as it should be, which coincides with here $\gamma_{\rm im}=\Im\gamma>0$ ($\gamma$ is the inverse scattering length).}
\begin{equation}\label{eq:1311}
{\rm Im} f(E)\geq \sqrt{2\mu E} |f(E)|^2, \qquad E>0,
\end{equation}
where $-\Im \kappa(E)=\sqrt{2\mu E}$ for $E>0$. Compared with Eq.~\eqref{eq:uni-multi}, one naturally obtains
$\gamma_{\rm im}>0$. The term $-\Im \kappa(E)$ is proportional to the elastic scattering cross-section; cf. Eq.~\eqref{eq:uni-elas}. Notably, the term $\Im f(E)$ also includes the contribution of the pole (or bound state) at $E_{\rm pole}=-\gamma^2/(2\mu)$ for $\gamma>0$; i.e., this pole
contributes a delta function to $\Im f(E)$. Physically, this can be attributed to the formation of the $X(3872)$ resonance.

The proximity of the $X(3872)$ mass and the $D^0\bar D^{*0}$ threshold requires the inclusion of the finite $D^{*0}$ width. This has been stressed in the previous analyses \cite{Braaten:2007ft,Kang:2016jxw}. Considering the isospin symmetry of strong coupling, the $D^{*0}$ width can be predicted
by the known $D^{*+}$ width. Below, we use the $D^{*0}$ width as $\Gamma_{*0}=(0.066\pm0.015)$ MeV. The variable $\kappa(E)$ will then be
$\kappa(E)=\sqrt{-2\mu (E+i\Gamma_{*0}/2)-i\epsilon}$. In the limits of $\gamma_{\rm im}\to 0$ and $\Gamma_{*0}\to 0$, we return to the simplest case.
Since we consider a width of $D^{*0}$, the ultimate final states are $D^0\bar D^0\pi^0$ and $D^0\bar D^0\gamma$, and $\bar D^0 D^+\pi^-$ and $D^0D^-\pi^+$
if $E>2.2$ MeV since $M_{D^{*0}}-M_{D^+}-M_{\pi^-}=-2.2$ MeV.

We obtain the scattering amplitude $f(E)$ as
\begin{eqnarray}\label{eq:fE}
f(E) = \frac{1}{-(\gamma_{\rm re}+i\gamma_{\rm im}) + \sqrt{-2\mu(E + i\Gamma_{*0}/2)}}.
\end{eqnarray}
This equation should be accurate within one MeV away from the $D^0\bar D^{*0}$ threshold (cf. the pion exchange scale of approximately 10 MeV mentioned in the introduction). The pole of $f(E)$ is $E_{\rm pole}=-E_X-i\Gamma_X/2$ with
\begin{eqnarray}
E_X &=& \frac{\gamma^2_{\rm re} - \gamma^2_{\rm im}}{2\mu},\\ \nonumber
\Gamma_X &=& \Gamma_{*0} +  \frac{2\gamma_{\rm re}\gamma_{\rm im}}{\mu}.
\end{eqnarray}
The residue at this pole is $-\gamma/\mu=-(\gamma_{\rm re}+i \gamma_{\rm im})/\mu$.
However, the definitions of the binding energy and width may be problematic. $E_X$ and $\Gamma_X$ can be understood as the binding energy and width only in the case of $X(3872)$ as a narrow resonance ($\Gamma_X\ll 2E_X$), whose line shape behaves like a (nonrelativistic) Breit--Wigner shape. If $\Gamma_X/(2E_X)$ is not small, this interpretation does not hold. Additionally, for the virtual state case ($\gamma_{\rm re}<0$), the variables $E_X$ and $\Gamma_X$ only specify a pole on the second Riemann sheet of complex energy $E$ but have no other precise physical interpretation.

The variables $E_X$ and $\Gamma_X$ cannot be directly measured in the experiment. They specify the pole position of $f(E)$ in the complex $E$ plane.
In the first Riemann sheet, we refer to the pole as the bound state in the sense of $\Gamma_X\to 0$, as is also done for $p\bar p$ scattering \cite{Kang:2016jxw}. In the second sheet we refer to it as the virtual state in the sense of $\Gamma_X\to 0$. Instead, the pair of variables $E_{\rm max}$ and $\Gamma_{\rm fwhm}$ are the observables in an experimental measurement for the line shape. The peak position $E_{\rm max}$ of $|f(E)|^2$ can be obtained by
\begin{equation}
2\mu E_{\rm max} + \gamma_{\rm re}\left(\mu\sqrt{E^2_{\rm max} + \Gamma^2_{*0}/4} - \mu E_{\rm max}\right)^{1/2} + \gamma_{\rm im}\left(\mu\sqrt{E^2_{\rm max} + \Gamma^2_{*0}/4} + \mu E_{\rm max}\right)^{1/2}= 0,
\end{equation}
and the full with at half maximum is given by $\Gamma_{\rm fwhm}=E_+-E_-$, with $E_{\pm}$ solved by the following equation:
\begin{equation}
\left| f(E_\pm) \right|^2=\frac{1}{2}\left| f(E_{\rm max}) \right|^2.
\end{equation}

For the decay mode of $J/\psi\pi^+\pi^-$, the energy dependence in the $X(3872)$ region derives solely from $|f(E)|^2$, which corresponds to the propagator of the $X(3872)$ resonance. We then have the line shape in the $J/\psi\pi^+\pi^-$ channel \cite{Braaten:2009jke}:
\begin{eqnarray}\label{eq:dGdE:Jpsipipi}
\frac{d\hat{\Gamma}_{\rm SD}}{dE} &=& \frac{\mu^2\Gamma_X}{2\pi(\gamma^2_{\rm re}+\gamma^2_{\rm im})}\left| f(E) \right|^2,
\end{eqnarray}
where ``SD'' means a short distance since, in this decay mode, $D^0$ and $\bar D^{*0}$ should approach each other within a small distance, unlike the constituent that decays to $D^0\bar D^{*0}$, where there is a large root-mean-square separation. The factor in front of $|f(E)|^2$ is chosen such that
one will have the correct normalization for a narrow resonance case. If $\Gamma_X \ll 2E_X$, the line shape in the region $[-2E_X, 0]$ is approximately a Breit--Wigner line shape, and the integral of $d\hat{\Gamma}_{\rm SD}/dE$ over this region is approximately 1.

For the decay mode of $D^0\bar D^0\pi^0$, the line shape is given by \cite{Braaten:2009jke}
\begin{equation}\label{eq:dGdEDDpi}
\frac{d\Gamma_{D^0\bar D^0\pi^0}}{dE}\propto |f(E)|^2\times \left(\sqrt{E^2+\Gamma_{*0}^2/4}+E\right)^{1/2}.
\end{equation}
This equation is obtained in Ref.~\cite{Braaten:2009jke} by considering the $D^*$ propagator and the three-body phase space integral.
As derived in Refs.~\cite{Braaten:2009jke} and \cite{Kang:2016jxw}, Eq.~\eqref{eq:dGdEDDpi} will be accurate once the signal occurs within approximately an MeV of the threshold. Reference \cite{Braaten:2007ft} shows a more elegant way to obtain the additional factor followed by $|f(E)|^2$ by noticing the expression of the optical theorem in Eq.~\eqref{eq:uni-multi}. The imaginary part, $-\Im \kappa(E)$, contributes to the $D^0\bar D^0\pi^0$ energy distribution. We can express the following complex variable by choosing the appropriate branching cut:
\begin{equation}
 \sqrt{-2\mu(E+i\Gamma_{*0}/2)}=\sqrt{\mu}\left[\left(\sqrt{E+\Gamma_{*0}^2/4}-E\right)^{1/2}-i \left(\sqrt{E+\Gamma_{*0}^2/4}+E\right)^{1/2}\right].
\end{equation}
By introducing the appropriate factor, we get
\begin{equation}\label{eq:dGdE-DDpi}
\frac{d\Gamma_{D^0\bar D^0\pi^0}}{dE}\propto\frac{d\hat\Gamma_{\rm SD}}{dE} \left(\frac{\sqrt{E^2+\Gamma_{*0}^2/4}+E}{\sqrt{E_X^2+\Gamma_{*0}^2/4}-E_X}\right)^{1/2}.
\end{equation}
At $E=-E_X$, the last factor decreases to 1. Equations \eqref{eq:dGdE:Jpsipipi} and \eqref{eq:dGdE-DDpi} are
as the line shapes for the $J/\psi\pi^+\pi^-$ and $D^0\bar D^0\pi^0$ channels respectively.

The line shapes for the $J/\psi\pi^+\pi^-$ and $D^0\bar D^0\pi^0$ channels differ for the bound state ($\gamma_{\rm re}>0$) and virtual state ($\gamma_{\rm re}<0$) cases if $X$ is sufficiently narrow. For the $J/\psi\pi^+\pi^-$ channel, if $X$ is a bound state, the line shape shows the Breit--Wigner shape below the $D^0\bar D^{*0}$ threshold, whereas there would be only a cusp at the $D^0\bar D^{*0}$ threshold for a virtual state. For the $D^0\bar D^0\pi^0$ channel, if $X$ is a bound state, the line shape shows a Breit--Wigner shape below the $D^0\bar D^{*0}$ threshold and a threshold enhancement above the $D^0\bar D^{*0}$ threshold.
However, there would be only a threshold enhancement above the $D^0\bar D^{*0}$ threshold for a virtual state. For an intuitive picture, see Figs.~2 and 3
in Ref.~\cite{Braaten:2007ft}. However, the increase in the width of $X$
would provide more smearing of the line shape. Therefore, the precise measurement of the $X$ width \cite{LHCb:2020xds,LHCb:2020fvo} in addition to mass measurement is crucial for determining the nature of the exotic $X(3872)$ state.

\section{FIT RESULTS and discussion}

In this section, we analyze recent data on the invariant mass distributions of $M(J/\psi\pi^+\pi^-)$ and $M(D^0\bar{D}^0\pi^0)$ for the $X(3872)$ resonance from the BESIII collaboration \cite{BESIII:2023hml}. We discover that those data can be rather well described in our analysis. When only the scattering length approximation of the scattering amplitude is used, a bound state or virtual state can be generated. In all the fits, $\gamma_{\rm re}$ takes positive values, indicating the interpretation of $X(3872)$ as a bound state. This constitutes our main conclusion on the nature of $X(3872)$. In terms of the concept of compositeness, $X(3872)$ can be interpreted as a pure bound state with a compositeness of 1 \cite{Kang:2016ezb,Kinugawa:2023fbf,Kang:2016jxw,Zhang:2022hfa,Wang:2022vga}.

\subsection{Fit results of $J/\psi\pi^+\pi^-$}\label{sec:Jpsipipi}
The background contribution is parameterized as
\begin{equation}
B(E) = aE + b,
\end{equation}
where $a$ and $b$ are the fitting parameters. We have examined several higher-order polynomials for parameterizing the background term, but they do not considerably improve the fit quality. In the region we considered, i.e., $-$100 MeV below the $D^0\bar D^{*0}$ threshold to 100 MeV above the threshold, the linear line is enough to describe the background.

To determine the mass resolution, the BESIII collaboration has performed a careful Monte Carlo study for both channels \cite{BESIII:2023hml}. As a result,
the resolution $R(E^{'},E)$ for $J/\psi\pi^+\pi^-$ is described by
\begin{equation}
R(E^{'},E)= \frac{1}{\sqrt{2\pi}\sigma}\exp{\left(-\frac{(E^{'}-(E+dE))^2}{2\sigma^2}\right)},
\end{equation}
where the mass shift is $dE = 0.96(E+m_{\rm th})-(3.52$\,MeV) and the energy-dependent mass resolution is $\sigma = 2.07(E+m_{\rm th})- (6.11$\,MeV). Note that we define $E$ as the energy relative to the $D^0\bar D^{*0}$ threshold.

The event number as a function of the $J/\psi\pi^+\pi^-$ invariant mass in an energy bin of width $\Delta = 5$ MeV centered at $E_i$ is written as
\begin{align}\label{eq:NJpsipipi}
N_i(E) &= \int^{E_i+\Delta/2}_{E_i-\Delta/2} dE^{'} \int^\infty_{-\infty} dE R(E^{'},E) \left[ (\mathcal{BB})_{J/\psi\pi^+\pi^-}\frac{d\hat{\Gamma}_{\rm SD}}{dE} + B(E) \right],
\end{align}
with the signal term given in Eq.~\eqref{eq:dGdE:Jpsipipi}. We have 5 parameters in total: $\gamma_{\rm re}$, $\gamma_{\rm im}$, $(\mathcal{BB})_{J/\psi\pi^+\pi^-}$, and $a$ and $b$ in the background term, which can be fitted to the experimental data.
For a narrow resonance case, where the typical line shape resembles a Breit-Winger one, the integration over $d\hat\Gamma_{\text SD}/dE$ is approximately 1. In this case, $(\mathcal{BB})_{J/\psi\pi^+\pi^-}$ can be interpreted as the corresponding experimental efficiency multiplied by the product of the event number for $e^+e^-\to \gamma X$ and the branching fraction of $X\to J/\psi\pi^+\pi^-$, and otherwise, it is just an overall normalization constant. $\mathcal{BB}$ corresponds to the terminology of ``yield'' used in our previous analysis \cite{Kang:2016jxw}.

\begin{figure}[htbp]
\centering
\includegraphics[width=4in]{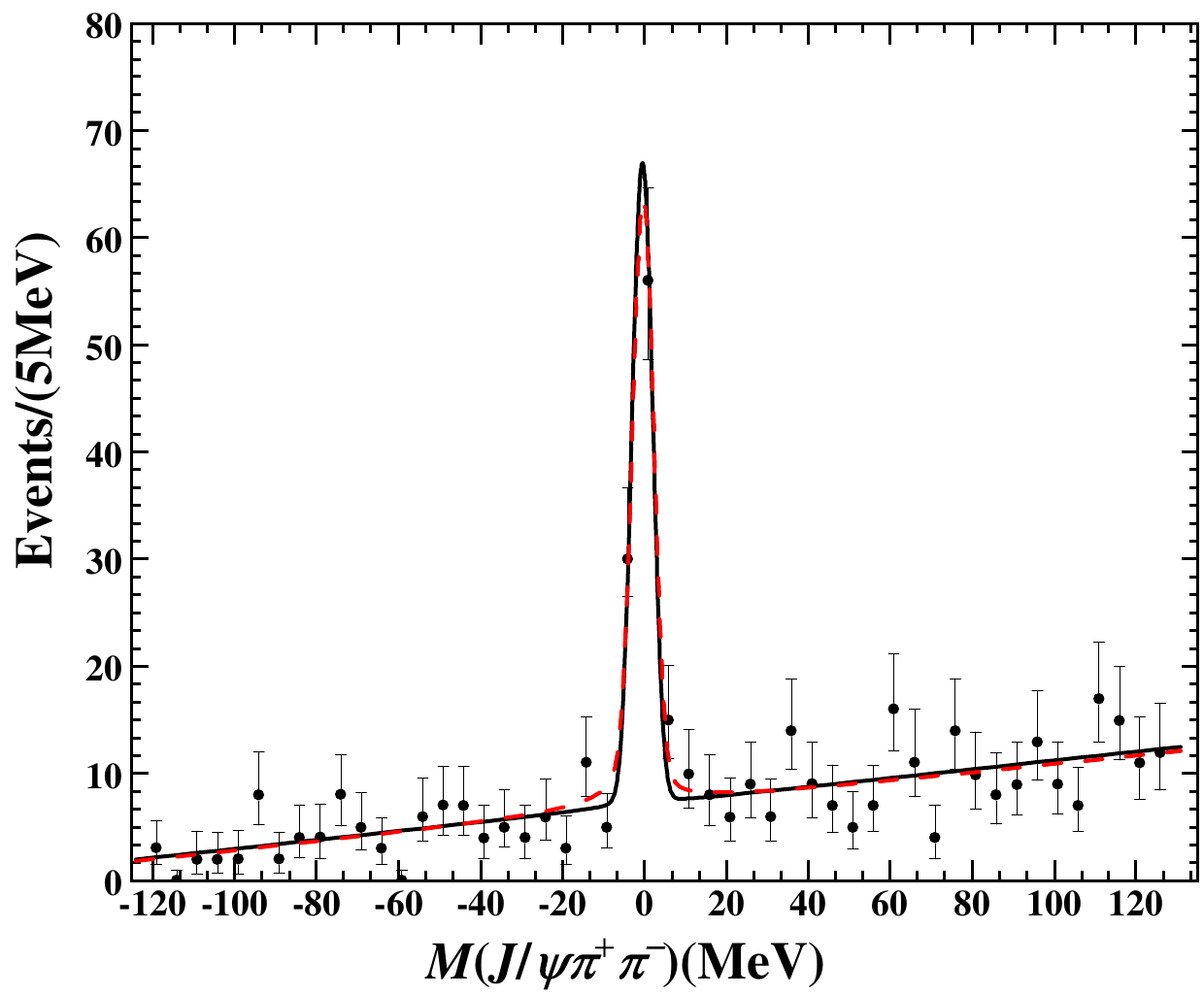}
\caption{Invariant mass distribution for the $J/\psi\pi^+\pi^-$ decay channel. The data are obtained from the BESIII collaboration \cite{BESIII:2023hml}. The solid line represents our fit result for $\gamma_{\rm re}+i\gamma_{\rm im}=35.0$ MeV. The dashed line represents our fit result for $\gamma_{\rm re}+i\gamma_{\rm im}=(22.6+i\,4.7)$ MeV.}
\label{fig:Jpsipipi}
\end{figure}

We use the maximum log(likelihood) method for the fit. The results are given in
Table~\ref{tab:Jpsipipi} and Fig.~\ref{fig:Jpsipipi}. The uncertainties for the parameters correspond to those produced by the MIGRAD method in the MINUIT package \cite{MINUIT}. The fits are rather good. We explore two types of fit: in one case, we leave $\gamma_{\rm im}$ free, and in the other case, we fix it to 0. Their fit qualities are essentially the same from the perspective of pragmatism, although the $-2$log(likelihood) of the $\gamma_{\rm im}\neq0$ case is smaller than that of $\gamma_{\rm im}=0$ by 1 from a purely academic viewpoint. Moreover, the uncertainty of $\gamma_{\rm im}$ is too large, which indicates that $\gamma_{\rm im}$ cannot be determined well by the data. In other words, $\gamma_{\rm im}=0$ works well enough for the current data. To reduce the number of parameters, we prefer to choose $\gamma_{\rm im}=0$. This choice is also adopted in the $D^0\bar D^0\pi^0$ analysis below as well as in the simultaneous fit. The best fit $\gamma_{\rm im}=0$ implies that the contribution of the inelastic channel $J/\psi\pi^+\pi^-$ to the width $\Gamma_X$ is negligible. This point seems to be favored by the current PDG \cite{ParticleDataGroup:2024cfk} values $\Gamma(X\to J/\psi\pi^+\pi^-)=(3.5 \pm 0.9)\%$, $\Gamma(X\to D^0\bar D^0\pi^0)=(45 \pm 21)\%$, $\Gamma(X\to D^0\bar D^{*0})=(34 \pm 12)\%$, i.e., the partial width of $J/\psi\pi^+\pi^-$ accounts for 1/10 of that of $D^0\bar D^0\pi^0$ (or $D^0\bar D^{*0}$), although there are large uncertainties. A direct measurement of BESIII collaboration yields a value of $\Gamma(X\to D^0\bar D^{*0})/\Gamma(X\to J/\psi\pi^+\pi^-)=11.77\pm3.09$ \cite{BESIII:2020nbj}.
One needs to be cautious in understanding the ratio of $\Gamma(X\to D^0\bar D^0\pi^0)/\Gamma(X\to J/\psi\pi^+\pi^-)<1.16$ at the 90\% confidence level reported in the same reference \cite{BESIII:2020nbj}, where $D^0\bar D^0\pi^0$ means the non$ D^0\bar D^{*0}$ three-body decay. This again demonstrates that $D^0\bar D^0\pi^0$ mostly derives from $D^0\bar D^{*0}$. Additionally, for the best fit parameter $\gamma_{\rm re}=35.0$ MeV, the condition $\Gamma_X\ll 2E_X$ is satisfied, and $(\mathcal{BB})_{J/\psi\pi^+\pi^-}$ can be regarded as the yield in the channel of $e^+e^-\to\gamma X\to \gamma(J/\psi\pi^+\pi^-)$.

The quantities $E_X$, $\Gamma_X$, and $E_{\rm max}$ and $\Gamma_{\rm fwhm}$ are also calculated, with the results shown in Table \ref{tab:Jpsipipi}.
Their central values are obtained from the central values of $\gamma_{\rm re}$ and $\gamma_{\rm im}$. The uncertainties are obtained by incorporating the uncertainties of $\gamma_{\rm re}$ and $\gamma_{\rm im}$. More explicitly, we discretize $\gamma_{\rm re}$ and $\gamma_{\rm im}$ (if not zero) into hundreds of values within one standard deviation and calculate the resulting quantities from these numbers. The maximum and minimum values are chosen. Thus, asymmetric uncertainties appear. For this bound state below threshold ($\gamma_{\rm re}>0$), $E_X$ and $\Gamma_X$ can be interpreted as the binding energy and width, respectively, and thus $E_{\rm max}\approx -E_X$ and $\Gamma_{\rm fwhm}\approx\Gamma_X$ within uncertainties. In the case of $\gamma_{\rm im}=0$, the $X$ width only derives from the $D^*$ width, and thus, $\Gamma_{\rm fwhm}=\Gamma_{*0}=\Gamma_X=(0.066\pm0.015)$ MeV.

\begin{table}[htbp]
    \centering
    \begin{tabular*}{0.9\columnwidth}{@{\extracolsep{\fill}}c|ccccc@{}}
        \hline\hline
         Parameters & $\gamma_{\rm re}$  &$\gamma_{\rm im}$  &$(\mathcal{BB})_{J/\psi\pi^+\pi^-}$  &$10^3 a$  &$b$\\
        \hline
         \multirow{2}{*}{Fit results} & $35.0 \pm 10.3$ & $0$ & $74.4\pm9.6$ & $8.3\pm1.0$ & $1.4\pm0.1$ \\
        & $22.6 \pm 17.4$ & $4.7\pm4.2$ & $70.2\pm15.6$ & $8.1\pm1.0$ & $1.3\pm0.1$ \\
        \hline
          \multirow{3}{*}{Calculated results}  &$-E_X$  &$\Gamma_X$ & $E_{\rm max}$    &$\Gamma_{\rm fwhm}$ &  \\
        \cline{2-6}
        & $-0.63^{+0.32}_{-0.43}$      &$0.066\pm0.015$          &$-0.63^{+0.32}_{-0.43}$   &$0.066\pm0.015$    & \\
        &$-0.25^{+0.26}_{-0.55}$       &$0.29^{+0.51}_{-0.21}$   &$-0.27^{+0.25}_{-0.56}$   &$0.29^{+0.51}_{-0.20}$ &\\
        \hline\hline
    \end{tabular*}
    \caption{Results of our data analysis for the $J/\psi\pi^+\pi^-$ mode from the BESIII collaboration \cite{BESIII:2023hml}. Both cases of fixing $\gamma_{\rm im} = 0$ and setting $\gamma_{\rm im}$ as a free parameter are shown. $\mathcal{BB}$ is dimensionless. $a$ is in units of MeV$^{-2}$, and $b$ is in units of MeV$^{-1}$. All others are in units of MeV.}
     \label{tab:Jpsipipi}
\end{table}

%%%%%%%%%%%%%%%%%%%%%%%%%%%%%%%%%%%%%%%%%%%%%%%%%%%%%%%%%%%%%%%%%%%%%%%%%%%%%%%%%%%%%%%%%%%%%%%%%%%%%%%%%%%%%%%%%%%%%%%%%%%%%%%%%%%%%%%%%%%%%%%%%%%%%%%%%%%%%%%%%%%%%%%%%%%%%%%%%%%%%%%%%%%%%%%%%%%%%%%%%%%%%%%%%%%%%%%%%%%%%%%%%%%%%%%%%%%%%%%%%%%%%%%%%%%%%%%%%%%%%%%%%%%%%%%%%%%%%%%%%%
\subsection{Fit results of $D^0\bar{D}^0\pi^0$}\label{sec:DDpifit}

The background contribution is parameterized as
\begin{equation}
B(E) = cE + d,
\end{equation}
where $c$ and $d$ are fitting parameters. Higher-order polynomials are also explored but do not significantly improve the fit quality. Similar to the above $J/\psi\pi^+\pi^-$ case, the experimental energy resolution $R(E^{'},E)$ is given by \cite{BESIII:2023hml}:
\begin{equation}
R(E^{'},E)= \frac{1}{\sqrt{2\pi}\sigma}\exp{\left(-\frac{(E^{'}-(E+dE))^2}{2\sigma^2}\right)},
\end{equation}
with the mass shift $dE = 0.092 $ MeV and $\sigma = 13.9(E+m_{\rm th}) - (53.0$\,MeV).
In Ref.~\cite{Braaten:2009jke}, the energy resolution used therein
is considered to be too crude an estimate, resulting in $\gamma_{\rm im}=0$ as an artifact; thus, the $D^0\bar D^{*0}$ analysis is only for illustrative purposes.
%Quality Control Editor: Please ensure that the intended meaning has been maintained in the following edit.
Here, the resolution function is carefully examined by the BESIII collaboration with a Monte Carlo method.

Another very important point is that the $D^0\bar D^0\pi^0$ data in Ref.~\cite{BESIII:2023hml} used here and those used in Refs.~\cite{Braaten:2009jke,Kang:2016jxw} should be understood differently. In the latter, the $D^0\bar D^0\pi^0$ event near the $D^0\bar D^{*0}$ threshold was assumed to derive from the $D^0\bar D^{*0}$ or $\bar D^0 D^{*0}$ pair, namely, a mass constraint from the mass of $D^{*0}$ was considered.
The former corresponds to a true $D^0\bar D^0\pi^0$ invariant mass distribution. Therefore, in Ref.~\cite{Braaten:2009jke}, Eq.~(22) for $d\Gamma/dE$ and Eq.~(25) for $d\Gamma/dE_{\rm exp}$ are different, and the authors also mentioned that the measurement of a true $D^0\bar D^0\pi^0$ spectrum is preferable to clarify whether consistent resonance parameters can be achieved compared with those from $J/\psi\pi^+\pi^-$.

\begin{figure}[htbp]
\centering

\includegraphics[width=4in]{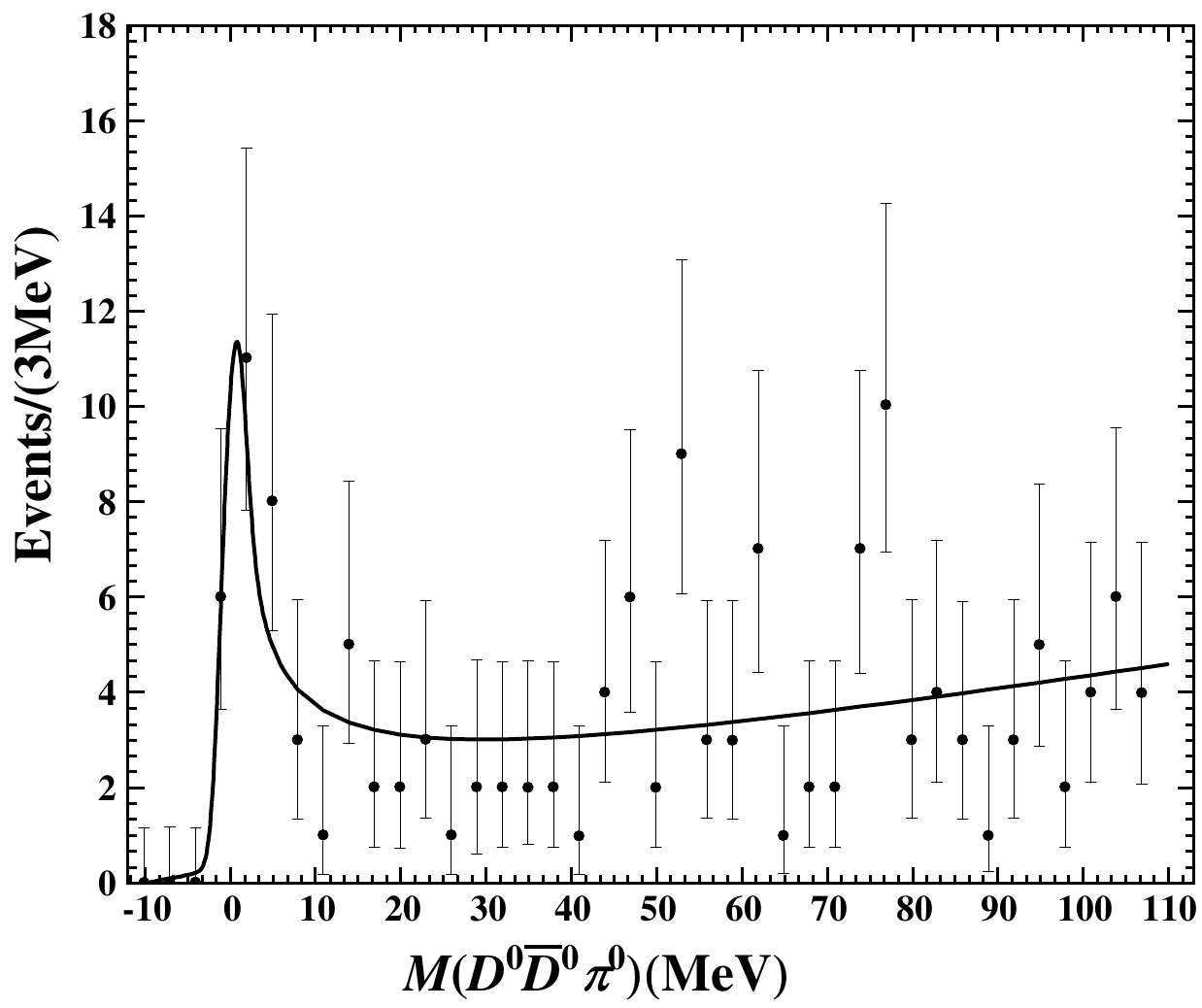}
\caption{Invariant mass distribution for the $D^0\bar{D}^0\pi^0$ decay channel. The data are obtained from the BESIII collaboration \cite{BESIII:2023hml}. The line represents our fit result.}
\label{fig:DDpi}
\end{figure}

The event number as a function of the $D^0\bar{D}^0\pi^0$ invariant mass in an energy bin of width $\Delta = 3$ MeV centered at $E_i$ can be written as
\begin{align}\label{eq:NDDpi}
N_i(E) &= \int^{E_i+\Delta/2}_{E_i-\Delta/2} dE^{'} \int^\infty_{-\infty} dE R(E^{'},E)\nonumber\\
 &\quad\times\left[ (\mathcal{BB})_{D^0\bar{D}^0\pi^0}\frac{d\hat\Gamma_{\rm SD}}{dE} \left(\frac{\sqrt{E^2+\Gamma_{*0}^2/4}+E}{\sqrt{E_X^2+\Gamma_{*0}^2/4}-E_X}\right)^{1/2} + B(E) \right].
\end{align}
We have 4 parameters in total: $\gamma_{\rm re}$, $(\mathcal{BB})_{D^0\bar{D}^0\pi^0}$, and the $c,\,d$ terms in the background.
In the case that the integration over $d\hat\Gamma_{\text SD}/dE$ is around 1, the parameter $(\mathcal{BB})_{D^0\bar D^0\pi^0}$ can be interpreted as the corresponding experimental efficiency multiplied by the product of the event number for $e^+e^-\to \gamma X$ and the branching fraction of $X\to D^0\bar D^0\pi^0$, and otherwise, it is just a convenient constant. $\gamma_{\rm im}$ has been fixed to 0 as the lowest possible value. We use the maximum log-likelihood method for the fit. The results are given in Fig.~\ref{fig:DDpi} and Table~\ref{tab:DDpi}. As shown in Fig.~\ref{fig:DDpi}, the experimental data are well reproduced with this simple model. We also explored the fit while leaving $\gamma_{\rm im}$ free, which prefers some negative values that can indeed maximize the $-2$log(likelihood). However, as we have mentioned, these values are unphysical, which violates the optical theorem. This situation agrees with the findings in Ref.~\cite{Braaten:2009jke}, where the best fit parameter of $\gamma_{\rm im}$ in Table II therein is not a positive value but 0 with a positive uncertainty. In Table~\ref{tab:DDpi}, we also find $E_{\rm max}\approx -E_X$.

\begin{table}[htbp]
    \centering
    \begin{tabular*}{0.9\columnwidth}{@{\extracolsep{\fill}}c|cccc@{}}
        \hline\hline
        Parameters & $\gamma_{\rm re}$  &$(\mathcal{BB})_{D^0\bar D^0\pi^0}$    &$c$             &$d$ \\
        Fit results & $11.3\pm9.4$      &$5.2\pm4.4$                            &$0.01\pm0.002$  &$0.1\pm0.02$ \\
        \hline
       \multirow{2}{*}{Calculated results} & $-E_X$  & $\Gamma_X$ & $E_{\rm max}$   &$\Gamma_{\rm fwhm}$  \\
         & $-0.07^{+0.06}_{-0.16}$   &$0.066\pm0.015$   & $-0.07^{+0.06}_{-0.15}$   &$0.066\pm0.015$   \\
        \hline\hline
    \end{tabular*}
    \caption{Results of our data analysis for the $D^0\bar{D}^0\pi^0$ decay channel from the BESIII collaboration \cite{BESIII:2023hml}. $\mathcal{BB}$ is dimensionless. $c$ is in units of MeV$^{-2}$, and $d$ is in units of MeV$^{-1}$. All others are in units of MeV.}
    \label{tab:DDpi}
\end{table}

%%%%%%%%%%%%%%%%%%%%%%%%%%%%%%%%%%%%%%%%%%%%%%%%%%%%%%%%%%%%%%%%%%%%%%%%%%%%%%%%%%%%%%%%%%%%%%%%%%%%%%%%%%%%%%%%%%%%%%%%%%%%%%%%%%%%%%%%%%%%%%%%%%%%%%%%%%%%%%%%%%%%%%%%%%%%%%%%%%%%%%%%%%%%%%%%%%%%%%%%%%%%%%%%%%%%%%%%%%%%%%%%%%%%%%%%%%%%%%%%%%%%%%%%%%%%%%%%%%%%%%%%%%%%%%%%%%%%%%%%%%
\subsection{A simultaneous fit}

\begin{figure}[htbp]
\centering

\includegraphics[width=6in]{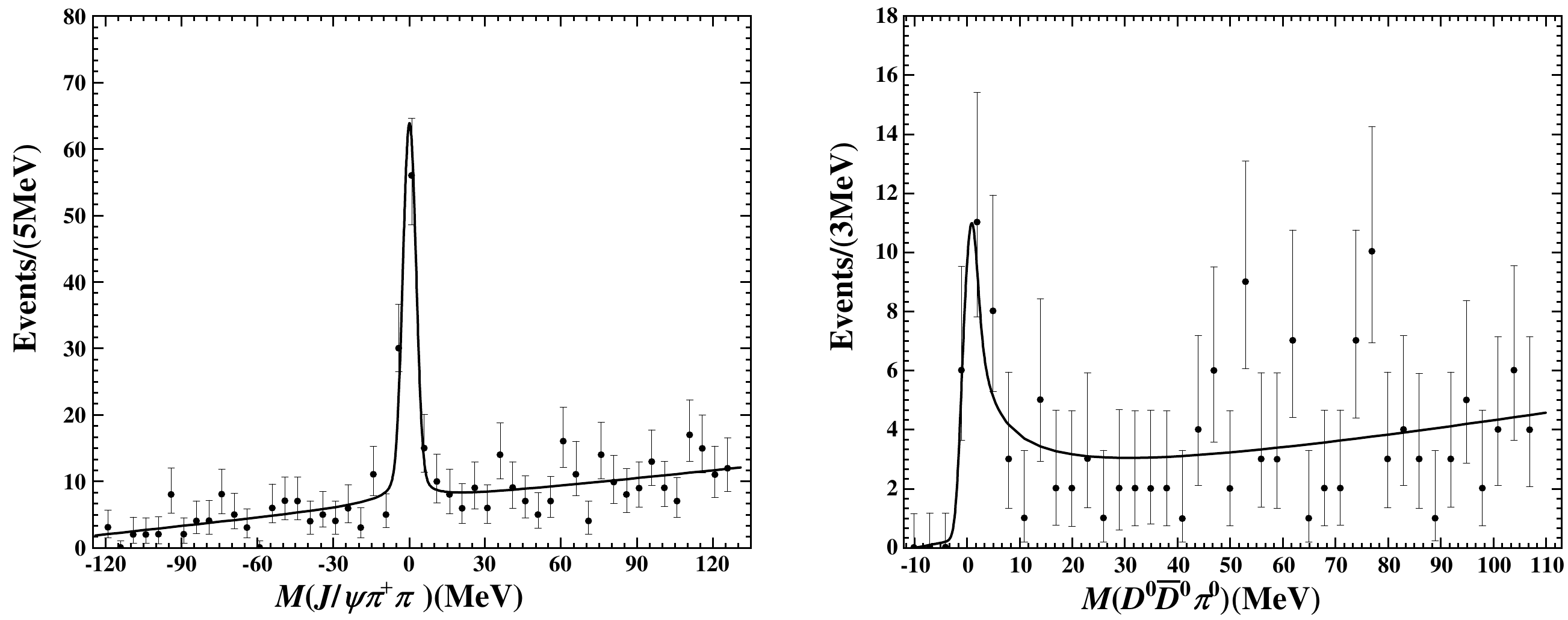}
\caption{Invariant mass distributions for the $J/\psi\pi^+\pi^-$ and $D^0\bar{D}^0\pi^0$ modes in a simultaneous fit. The data are obtained from the BESIII collaboration \cite{BESIII:2023hml}. The lines are our fit results.}
\label{fig:combined}
\end{figure}

In this section, we perform a simultaneous fit with the inclusion of the two datasets above. In this way, the parameters, e.g., $\gamma_{\rm re}$ specifying the pole position of the $X$ resonance, can be more constrained. We have 7 parameters in total: $\gamma_{\rm re}$, $(\mathcal{BB})_{J/\psi\pi^+\pi^-}$, $(\mathcal{BB})_{D^0\bar{D}^0\pi^0}$, and the $a,\,b,\,c,\,d$ terms in the background. $\gamma_{\rm im}$ has been fixed to 0, as discussed in Sec.~\ref{sec:DDpifit}. We use the maximum log-likelihood method for the fit. The results are given in Fig.~\ref{fig:combined} and Table~\ref{tab:combined}, which are our preferred results compared with those in Secs.~\ref{sec:Jpsipipi} and \ref{sec:DDpifit}. The experimental data are reproduced rather well. Again, we find that $E_{\rm max}\approx -E_X$ and $\Gamma_{\rm fwhm}=\Gamma_{*0}=\Gamma_X=(0.066\pm0.015)$ MeV because $\gamma_{\rm im}=0$. The preferred $\gamma_{\rm re}$ takes a value of $(8.2\pm 5.6)$ MeV, indicating a very loosely bound state with a binding energy of 0.03 MeV. The corresponding scattering length is $(29.5\pm14.8)$ fm. This is really a large scattering length (recalling the scattering length of 5.4 fm for nucleon--nucleon scattering in the $^3S_1$ wave), which justifies the applicability of the current approach. Since $\Gamma_{X}\ll 2E_X$ is not fulfilled, $(\mathcal{BB})_{J/\psi\pi^+\pi^-}$ and $(\mathcal{BB})_{D^0\bar D^0\pi^0}$ cannot be understood as yields. They are only appropriate normalization constants; thus, we cannot obtain more information from their ratio.

\begin{table}[htbp]
    \centering
    \begin{tabular*}{0.9\columnwidth}{@{\extracolsep{\fill}}c|cccc@{}}
        \hline\hline
        Parameters & $\gamma_{\rm re}$   &$(\mathcal{BB})_{J/\psi\pi^+\pi^-}$ &  $(\mathcal{BB})_{D^0\bar{D}^0\pi^0}$  & \\
        Fit results & $8.2\pm5.6$        &$43.0\pm 31.0$            &$3.6\pm3.0$    &  \\
        \hline
        Parameters & $10^3a$ & $b$ & $c$ & $d$ \\
        Fit results & $8.0\pm1.0$  & $1.3\pm0.09$ & $0.01\pm0.002$ & $0.1\pm0.02$ \\
        \hline
        \multirow{2}{*}{Calculated results} & $-E_X$  & $\Gamma_X$ & $E_{\rm max}$ &  $\Gamma_{\rm fwhm}$   \\
          & $-0.03^{+0.03}_{-0.03}$    &$0.066\pm0.015$     &$-0.04^{+0.03}_{-0.06}$   &$0.066\pm0.015$    \\
        \hline\hline
    \end{tabular*}
    \caption{Results of our analysis by considering the data for both $J/\psi\pi^+\pi^-$ and $D^0\bar{D}^0\pi^0$ from the BESIII collaboration \cite{BESIII:2023hml}. $(\mathcal{BB})$s are dimensionless. $a$ and $c$ are in units of MeV$^{-2}$, and $b$ and $d$ are in units of MeV$^{-1}$.
    All others are in units of MeV.}
    \label{tab:combined}
\end{table}

\section{Summary}
In 2023, the BESIII collaboration published new data on the $X(3872)$ state from $J/\psi\pi^+\pi^-$ and $D^0\bar{D}^0\pi^0$ invariant mass distributions. This stimulated our present analysis by considering the universal feature for an $S$-wave threshold resonance, where only the scattering length term is kept in the effective range expansion. To consider the inelastic effect due to channels such as $J/\psi\pi^+\pi^-$ and the finite $D^{*0}$ width effect, we extend the traditional scattering amplitude $f(E)$ to the form of Eq.~\eqref{eq:fE}.
By considering the appropriate normalization condition, we obtain the line shapes in Eqs.~\eqref{eq:dGdE:Jpsipipi} and ~\eqref{eq:dGdE-DDpi} for $J/\psi\pi^+\pi^-$ and $D^0\bar{D}^0\pi^0$, respectively. Considering the experimental resolution and background, we obtain our final theoretical formula Eqs.~\eqref{eq:NJpsipipi} and \eqref{eq:NDDpi} for describing the data. We consider the separate fits for $J/\psi\pi^+\pi^-$ and $D^0\bar{D}^0\pi^0$
as well as the simultaneous fit which provides a better constraint for the parameters. For the simultaneous fit, we show the results in Fig.~\ref{fig:combined} and Table~\ref{tab:combined}. The best fit value is $\gamma_{\rm re}=(8.2\pm5.6)$ MeV, and the resulting pole position is
$(-0.03-i0.03)$ MeV with the width solely deriving from the constituent width $D^{*0}$. We conclude that the new BESIII data imply that $X(3872)$ is a loosely bound state of $D^0\bar D^{*0}$ or is called by a charm meson molecule.

As an extension, we can include the couple-channel effect from $D^+\bar D^{*-}$ in the future, namely, considering the scattering between neutral $D^0\bar D^{*0}$ and charged $D^+\bar D^{*-}$, although the latter is far from the former by 8 MeV. By doing so, we can describe the line shape over a larger energy region. In addition, many new results may be obtained once these effects are included. For example, in Ref.~\cite{Kang:2016jxw}, a scenario of the simultaneous bound and virtual state appearing in the adjacent second and third sheets is obtained, which indicates a large portion of the $c\bar c$ component in the configuration of $X(3872)$.

\section*{Acknowledgments}
We gratefully acknowledge the helpful discussions with Dr.~Meng-Chuan Du in IHEP, China, regarding various experimental details. This work is supported by the National Natural Science Foundation of China under Project No. 12275023.


\begin{thebibliography}{99}

\bibitem{Belle:2003nnu}
S.~K.~Choi \textit{et al.} [Belle],
%``Observation of a narrow charmonium-like state in exclusive $B^\pm \to K^\pm \pi^+ \pi^- J/\psi$ decays,''
Phys. Rev. Lett. \textbf{91}, 262001 (2003)
[arXiv:hep-ex/0309032 [hep-ex]].
%2594 citations counted in INSPIRE as of 02 Oct 2024


\bibitem{Guo:2017jvc}
F.~K.~Guo, C.~Hanhart, U.~G.~Mei\ss{}ner, Q.~Wang, Q.~Zhao and B.~S.~Zou,
%``Hadronic molecules,''
Rev. Mod. Phys. \textbf{90}, no.1, 015004 (2018)
[erratum: Rev. Mod. Phys. \textbf{94}, no.2, 029901 (2022)]
[arXiv:1705.00141 [hep-ph]].


\bibitem{Brambilla:2019esw}
N.~Brambilla, S.~Eidelman, C.~Hanhart, A.~Nefediev, C.~P.~Shen, C.~E.~Thomas, A.~Vairo and C.~Z.~Yuan,
%``The $XYZ$ states: experimental and theoretical status and perspectives,''
Phys. Rept. \textbf{873}, 1-154 (2020)
[arXiv:1907.07583 [hep-ex]].


\bibitem{Chen:2022asf}
H.~X.~Chen, W.~Chen, X.~Liu, Y.~R.~Liu and S.~L.~Zhu,
%``An updated review of the new hadron states,''
Rept. Prog. Phys. \textbf{86}, no.2, 026201 (2023)
[arXiv:2204.02649 [hep-ph]].

\bibitem{Meng:2022ozq}
L.~Meng, B.~Wang, G.~J.~Wang and S.~L.~Zhu,
%``Chiral perturbation theory for heavy hadrons and chiral effective field theory for heavy hadronic molecules,''
Phys. Rept. \textbf{1019}, 1-149 (2023)
[arXiv:2204.08716 [hep-ph]].


\bibitem{Liu:2024uxn}
M.~Z.~Liu, Y.~W.~Pan, Z.~W.~Liu, T.~W.~Wu, J.~X.~Lu and L.~S.~Geng,
%``Three ways to decipher the nature of exotic hadrons: Multiplets, three-body hadronic molecules, and correlation functions,''
Phys. Rept. \textbf{1108}, 1-108 (2025)
[arXiv:2404.06399 [hep-ph]].


\bibitem{Kang:2016jxw}
X.~W.~Kang and J.~A.~Oller,
%``Different pole structures in line shapes of the $X(3872)$,''
Eur. Phys. J. C \textbf{77}, no.6, 399 (2017)
[arXiv:1612.08420 [hep-ph]].

\bibitem{Wang:2023sii}
Z.~G.~Wang,
%``Decipher the width of the X(3872) via the QCD sum rules,''
Phys. Rev. D \textbf{109}, no.1, 014017 (2024)
[arXiv:2310.02030 [hep-ph]].

\bibitem{Dias:2024zfh}
J.~M.~Dias, T.~Ji, X.~K.~Dong, F.~K.~Guo, C.~Hanhart, U.~G.~Mei\ss{}ner, Y.~Zhang and Z.~H.~Zhang,
%``A model-independent analysis of the isospin breaking in the $X(3872)~\to~J/\psi \pi^+\pi^-$ and $X(3872)~\to~J/\psi \pi^+\pi^0\pi^-$ decays,''
[arXiv:2409.13245 [hep-ph]].
%0 citations counted in INSPIRE as of 02 Oct 2024

\bibitem{Achasov:2024ezv}
N.~N.~Achasov and G.~N.~Shestakov,
%``Toward an estimate of the amplitude X(3872)\textrightarrow{}\ensuremath{\pi}0\ensuremath{\chi}c1(1P),''
Phys. Rev. D \textbf{109}, no.3, 036028 (2024)
[arXiv:2401.04948 [hep-ph]].


\bibitem{Yu:2023nxk}
S.~Y.~Yu and X.~W.~Kang,
%``Nature of X(3872) from its radiative decay,''
Phys. Lett. B \textbf{848}, 138404 (2024)
[arXiv:2308.10219 [hep-ph]].


\bibitem{BESIII:2024ync}
M.~Ablikim \textit{et al.} [BESIII],
%``Search for the radiative transition \ensuremath{\chi}c1(3872)\textrightarrow{}\ensuremath{\gamma}\ensuremath{\psi}2(3823),''
Phys. Rev. D \textbf{110}, no.1, 012012 (2024)
[arXiv:2405.07741 [hep-ex]].

\bibitem{BESIII:2023hml}
M.~Ablikim \textit{et al.} [BESIII],
%``Coupled-Channel Analysis of the \ensuremath{\chi}c1(3872) Line Shape with BESIII Data,''
Phys. Rev. Lett. \textbf{132}, no.15, 151903 (2024)
[arXiv:2309.01502 [hep-ex]].


\bibitem{ParticleDataGroup:2024cfk}
S.~Navas \textit{et al.} [Particle Data Group],
%``Review of particle physics,''
Phys. Rev. D \textbf{110}, no.3, 030001 (2024)

\bibitem{Braaten:2007dw}
E.~Braaten and M.~Lu,
%``Line shapes of the X(3872),''
Phys. Rev. D \textbf{76}, 094028 (2007)
[arXiv:0709.2697 [hep-ph]].

\bibitem{Braaten:2007ft}
E.~Braaten and M.~Lu,
%``The Effects of charged charm mesons on the line shapes of the X(3872),''
Phys. Rev. D \textbf{77}, 014029 (2008)
[arXiv:0710.5482 [hep-ph]].

\bibitem{Braaten:2009jke}
E.~Braaten and J.~Stapleton,
%``Analysis of J/psi pi+ pi- and D0 anti-D0 pi0 Decays of the X(3872),''
Phys. Rev. D \textbf{81}, 014019 (2010)
[arXiv:0907.3167 [hep-ph]].

\bibitem{Song:2023pdq}
J.~Song, L.~R.~Dai and E.~Oset,
%``Evolution of compact states to molecular ones with coupled channels: The case of the X(3872),''
Phys. Rev. D \textbf{108}, no.11, 114017 (2023)
[arXiv:2307.02382 [hep-ph]].

\bibitem{Fleming:2007rp}
S.~Fleming, M.~Kusunoki, T.~Mehen and U.~van Kolck,
%``Pion interactions in the $X(3872)$,''
Phys. Rev. D \textbf{76}, 034006 (2007)
[arXiv:hep-ph/0703168 [hep-ph]].

\bibitem{Kang:2013uia}
X.~W.~Kang, J.~Haidenbauer and U.~G.~Mei\ss{}ner,
%``Antinucleon-nucleon interaction in chiral effective field theory,''
JHEP \textbf{02}, 113 (2014)
[arXiv:1311.1658 [hep-ph]].

\bibitem{Kang:2015yka}
X.~W.~Kang, J.~Haidenbauer and U.~G.~Mei\ss{}ner,
%``Near-threshold $\bar pp$ invariant mass spectrum measured in $J/\psi$ and $\psi'$ decays,''
Phys. Rev. D \textbf{91}, no.7, 074003 (2015)
[arXiv:1502.00880 [nucl-th]].


\bibitem{Haidenbauer:2015yka}
J.~Haidenbauer, C.~Hanhart, X.~W.~Kang and U.~G.~Mei\ss{}ner,
%``Origin of the structures observed in $e^+e^-$ annihilation into multipion states around the $\bar pp$ threshold,''
Phys. Rev. D \textbf{92}, no.5, 054032 (2015)
[arXiv:1506.08120 [nucl-th]].


\bibitem{LHCb:2020xds}
R.~Aaij \textit{et al.} [LHCb],
%``Study of the lineshape of the $\chi_{c1}(3872)$ state,''
Phys. Rev. D \textbf{102}, no.9, 092005 (2020)
[arXiv:2005.13419 [hep-ex]].

\bibitem{LHCb:2020fvo}
R.~Aaij \textit{et al.} [LHCb],
%``Study of the $\psi_2(3823)$ and $\chi_{c1}(3872)$ states in $B^+ \rightarrow \left( J\psi\pi^+\pi^-\right)K^+$ decays,''
JHEP \textbf{08}, 123 (2020)
[arXiv:2005.13422 [hep-ex]].

\bibitem{Kang:2016ezb}
X.~W.~Kang, Z.~H.~Guo and J.~A.~Oller,
%``General considerations on the nature of $Z_b(10610)$ and $Z_b(10650)$ from their pole positions,''
Phys. Rev. D \textbf{94}, no.1, 014012 (2016)
[arXiv:1603.05546 [hep-ph]].


\bibitem{Zhang:2022hfa}
L.~Zhang, X.~W.~Kang and X.~H.~Guo,
%``Composite nature of $Z_b$ states from data analysis,''
Eur. Phys. J. C \textbf{82}, no.4, 375 (2022)
[arXiv:2203.02301 [hep-ph]].


\bibitem{Kinugawa:2023fbf}
T.~Kinugawa and T.~Hyodo,
%``Compositeness of Tcc and X(3872) by considering decay and coupled-channels effects,''
Phys. Rev. C \textbf{109}, no.4, 045205 (2024)
[arXiv:2303.07038 [hep-ph]].


\bibitem{Wang:2022vga}
Z.~Q.~Wang, X.~W.~Kang, J.~A.~Oller and L.~Zhang,
%``Analysis on the composite nature of the light scalar mesons f0(980) and a0(980),''
Phys. Rev. D \textbf{105}, no.7, 074016 (2022)
[arXiv:2201.00492 [hep-ph]].


\bibitem{MINUIT}
F.~James,``MINUIT ¨C Function Minimization and Error Analysis'', CERN Program Library Long
Writeup D506, Version 94.1.


\bibitem{BESIII:2020nbj}
M.~Ablikim \textit{et al.} [BESIII],
%``Study of Open-Charm Decays and Radiative Transitions of the $X(3872)$,''
Phys. Rev. Lett. \textbf{124}, no.24, 242001 (2020)
[arXiv:2001.01156 [hep-ex]].
%44 citations counted in INSPIRE as of 04 Oct 2024
\end{thebibliography}
\end{document}